\documentclass{bmvc2k}

\pdfoutput=1
\title{A Learnable Color Correction Matrix for RAW Reconstruction}

\newcommand{\firstauthor}{Anqi Liu\textsuperscript{1}}
\newcommand{\secondauthor}{Shiyi Mu\textsuperscript{2}}
\newcommand{\correspondingauthor}{Shugong Xu\textsuperscript{*}}

\addauthor{\firstauthor}{liuanqiqiqi@shu.edu.cn}{1}
\addauthor{\secondauthor}{shiyimu@shu.edu.cn}{2}
\addauthor{\correspondingauthor}{shugong@shu.edu.cn}{*}

\addinstitution{
 School of Communication and\\
 Information Engineering,\\
 Shanghai University,\\
 Shanghai, China
}

\runninghead{LIU, MU, XU}{A Learnable Color Correction Matrix for RAW Reconstruction}

\usepackage{multirow}
\usepackage{hyperref}
\usepackage{fontawesome}

\begin{document}

\maketitle

\begin{abstract}
Autonomous driving algorithms usually employ sRGB images as model input due to their compatibility with the human visual system. However, visually pleasing sRGB images are possibly sub-optimal for downstream tasks when compared to RAW images. The availability of RAW images is constrained by the difficulties in collecting real-world driving data and the associated challenges of annotation. To address this limitation and support research in RAW-domain driving perception, we design a novel and ultra-lightweight RAW reconstruction method. The proposed model introduces a learnable color correction matrix (CCM), which uses only a single convolutional layer to approximate the complex inverse image signal processor (ISP). Experimental results demonstrate that simulated RAW (simRAW) images generated by our method provide performance improvements equivalent to those produced by more complex inverse ISP methods when pretraining RAW-domain object detectors, which highlights the effectiveness and practicality of our approach.
\end{abstract}

%-------------------------------------------------------------------------
\section{Introduction}
\label{sec:intro}
The construction of datasets is crucial for autonomous driving algorithms, as a large volume of data is needed to train models for recognizing and understanding road environments. Existing widely used real-world autonomous driving datasets, such as \cite{caesar2020nuscenes, sun2020scalability, cordts2016cityscapes, chang2019argoverse, huang2019apolloscape, yu2020bdd100k}, only provide sRGB images that are derived from RAW images through a camera image signal processor (ISP). Unlike sRGB images, RAW images offer physically meaningful and interpretable information due to their linear relationship between image intensity and the radiant energy incident on the camera sensor. This characteristic of RAW images has been effectively utilized in low-level tasks such as denoising \cite{mildenhall2018burst, schwartz2018deepisp, abdelhamed2018high, wei2020physics, chen2019seeing}, deblurring \cite{rim2020real}, HDR \cite{liu2020single, zou2023rawhdr, chen2018learning, mildenhall2022nerf}, and super-resolution \cite{tang2022learning, zhang2019zoom}. Additionally, RAW images have demonstrated advantages in high-level tasks such as object detection \cite{xu2023toward, hong2021crafting, li2024efficient, chan2023raw, sasagawa2020yolo, morawski2022genisp} and segmentation \cite{chen2023instance}.

Due to the suitability with the human visual system and the abundance and diversity of sRGB images, most autonomous driving algorithms use sRGB images as input. However, sRGB images may not always be the most suitable data type for computer learning compared to RAW images. As demonstrated by Tesla \cite{tesla2022ai}, using 12-bit RAW photon count images as input to the network bypasses the complex process of manually tuning the ISP and improves algorithm processing speed. Existing research on RAW-domain object detection \cite{ljungbergh2023raw, xu2023toward, hong2021crafting, li2024efficient, chan2023raw} shows that using RAW images as model input can lead to better performance.

However, the availability of RAW images is limited due to the constraints of data collection and the high cost of RAW image transmission, storage, and processing. Additionally, annotating RAW images is expensive and burdensome. As a result, only a few RAW datasets \cite{xu2023toward, hong2021crafting, omid2014pascalraw} of real-world driving scenarios are available. One solution to expand the availability of RAW images is to reconstruct them from existing sRGB images using inverse ISP algorithms. Traditional methods \cite{brooks2019unprocessing, conde2022model, koskinen2019reverse} gradually convert sRGB images to RAW images by simulating ISP system modules. Metadata-based methods \cite{li2023metadata, nam2022learning, punnappurath2021spatially, wang2023beyond, wang2023raw} encode necessary metadata from RAW images into the inverse ISP model to achieve high-accuracy reconstruction. Learning-based methods \cite{xing2021invertible, zamir2020cycleisp, afifi2021cie, otsuka2023self, kinli2022reversing, zhou2021raw, nam2022learning, nguyen2016raw} aim to approximate the mapping from sRGB images to RAW images using neural networks in an end-to-end pipeline. Most of these aforementioned methods \cite{xing2021invertible, zamir2020cycleisp, afifi2021cie, conde2022model, koskinen2019reverse} require paired sRGB and RAW images for training, which limits their application to converting existing sRGB datasets captured by unknown cameras into RAW format. To address this limitation, \cite{li2024efficient} propose an unpaired RAW reconstruction pipeline based on CycleGAN \cite{zhu2017unpaired}, which can convert sRGB images to RAW images for any camera style, making it ideal for performing RAW reconstruction in the absence of sRGB and RAW image pairs.

In practical industrial applications, enterprises have accumulated extensive sRGB images and high-quality labels through long-term real-world road tests and simulation trials, which are used for iterative optimization of autonomous driving algorithms. These sRGB images are invaluable for RAW-domain driving perception tasks, as they can be converted into RAW format using inverse ISP algorithms. Additionally, the labels associated with these images can be directly reused for subsequent tasks, eliminating the need for costly relabeling. However, converting large volumes of sRGB images into RAW format is time-consuming and computationally expensive. To advance RAW-domain research, there is a pressing need for a time-efficient RAW reconstruction method with high reconstruction quality. To address this challenge, we propose a novel learning-based method for RAW reconstruction. Our proposed model is exceptionally simple and lightweight. Specifically, the complex inverse ISP system is replaced with a simple and learnable color correction matrix, implemented using just a single convolutional layer. Our contributions are summarized as follows:
\begin{itemize}
  \item We propose a novel inverse ISP method for RAW reconstruction. Our proposed model consists of only one convolution layer to learn the mapping from sRGB images to RAW images.
  \item In the absence of real-world driving scene RAW datasets, our proposed model can rapidly expand the quantity of available RAW images, making it highly practical for advancing research in RAW-domain autonomous driving algorithms.
  \item Experimental results show that when used to pretrain a RAW-domain object detector, the simulated RAW (simRAW) images generated by our learnable color correction matrix achieve equivalent performance gains comparable to those generated by more complex inverse ISP methods. This demonstrates the feasibility and effectiveness of our proposed approach.
\end{itemize}

\section{Related Works}

\begin{table}
\begin{center}
\begin{tabular}{c|c|c|c|c|c}
\hline
Dataset                   & Sensor           & Bit & Images & Scenario    & Task    \\ \hline
PASCALRAW \cite{omid2014pascalraw} & Nikon D3200 DSLR & 12        & 4259   & Day         & Det     \\ \hline
ROD \cite{xu2023toward}   & Sony IMX490      & 24        & 25207  & Day\&Night  & Det     \\ \hline
\multirow{5}{*}{multiRAW \cite{li2024efficient}} & huawei P30pro   & 12        & 3004   & Day\&Night  & Det     \\ \cline{2-6} 
                          & iPhone XSmax    & 12        & 1153   & Day         & Det,Seg \\ \cline{2-6} 
                          & asi 294mcpro    & 14        & 2950   & Day\&Night  & Det     \\ \cline{2-6} 
                          & Oneplus 5t      & 10        & 101    & Day         & Det     \\ \cline{2-6} 
                          & LUCID TRI054S    & 24        & 261    & Day\&Tunnel & Det     \\ \hline
\end{tabular}
\end{center}
\caption{RAW datasets of real-world driving scenarios. The \textit{Sensor} indicates the camera sensor used to collect the dataset. The \textit{Bit} indicates the bit depth used to represent each pixel in the digital image, with higher bit depths providing better image quality and more accurate color and shade reproduction. The \textit{Det} indicates object detection tasks, the \textit{Seg} indicates segmentation tasks.}
\label{tab:dataset}
\end{table}

{\bf RAW datasets of driving scenarios.} To our knowledge, only a few RAW datasets are available as shown in Table \ref{tab:dataset}. The PASCALRAW dataset \cite{omid2014pascalraw} is collected using a Nikon D3200 DSLR camera and contains 4259 high-resolution 12-bit RAW images of daytime scenes. The ROD dataset \cite{xu2023toward} is collected using the Sony IMX490 sensor and contains 25k high-resolution 24-bit RAW images of both daytime and nighttime scenarios. The multiRAW dataset \cite{li2024efficient} is collected using five different cameras and contains 7469 high-bit RAW images from various times and locations (rural, tunnel, and urban areas), with corresponding sRGB images converted using the respective camera ISP are provided.

\noindent {\bf RAW Reconstruction.} Traditional inverse ISP methods \cite{brooks2019unprocessing} typically involve non-learnable operations based on camera parameters, such as inverse tone mapping, inverse gamma transformation, inverse color correction, inverse white balance and mosaic. This paradigm requires manually setting the true internal camera parameters which are often proprietary and not publicly available. Additionally, they do not account for complex nonlinear ISP operations, leading to inaccurate reconstruction of sRGB and RAW images. Metadata-based methods \cite{li2023metadata, nam2022learning, punnappurath2021spatially, wang2023beyond, wang2023raw} incorporate necessary metadata of RAW images in the network to guide RAW reconstruction process. Studies such as \cite{zamir2020cycleisp, xing2021invertible, kim2023paramisp} focus on learning-based methods, where ISP operations are learned end-to-end by neural networks. For instance, CycleISP \cite{zamir2020cycleisp} uses cycle consistency to learn both the forward RAW-to-sRGB and reverse sRGB-to-RAW translations. InvISP \cite{xing2021invertible} proposes an flow-based invertible ISP architecture using a single invertible neural network for both forward and inverse translations. ParamISP \cite{kim2023paramisp} presents a hybrid model that combines model-based and data-driven approaches, leveraging standard ISP operations in a learnable and interpretable manner. In this approach, camera parameters from EXIF data are converted into feature vectors to control the ISP network. Additionally, \cite{li2024efficient} introduces an unpaired RAW-to-sRGB ISP and sRGB-to-RAW inverse ISP based on CycleGAN \cite{zhu2017unpaired}, which eliminates the need for paired data and is more accessible. While existing methods achieve high-quality reconstruction, there remains a need for a more concise and equally accurate inverse ISP method to enable high-speed RAW reconstruction for practical applications, as discussed in Sec~\ref{sec:intro}.

\noindent {\bf RAW-domain Object Detection}. Existing research on RAW-domain object detection \cite{ljungbergh2023raw, xu2023toward, hong2021crafting, li2024efficient, chan2023raw} indicates that using RAW images as model input can yield better performance compared to sRGB images. For instance, \cite{xu2023toward} proposes a RAW adapter that is image-adaptive and jointly optimized with downstream detectors. \cite{ljungbergh2023raw} argues that ISPs should be optimized for specific tasks, suggesting that Bayer pattern RAW images can be fed directly into detectors after a simple module such as Yeo-Johnson transformation. \cite{li2024efficient} indicates that detector performance can be improved by using RAW images with gamma correction. In these approaches, an adjustment module is usually employed to modify the data distribution of RAW images and enhance feature visibility.

\section{Method}

\begin{figure*}
\includegraphics[width=\linewidth]{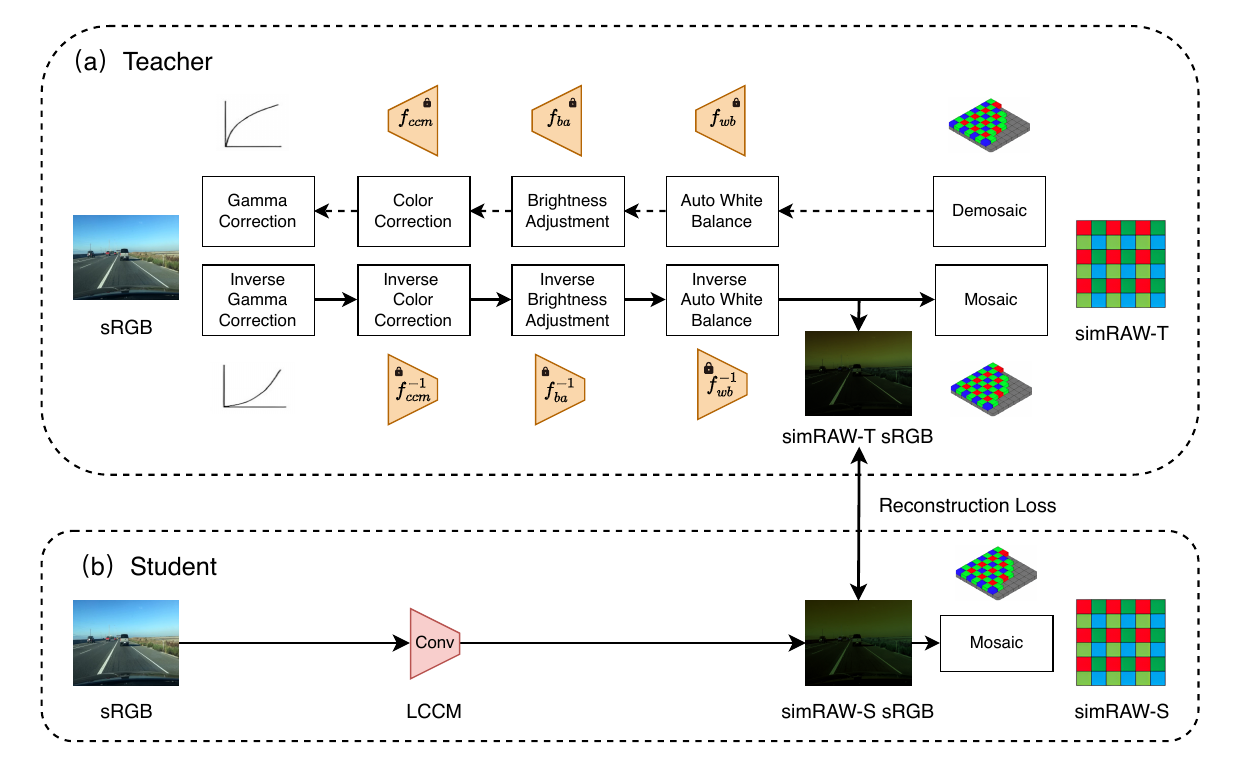}
    \caption{Overview of our proposed framework. The Unpaired-CycleR2R \cite{li2024efficient} model serves as the teacher model, while our proposed LCCM functions as the student model. LCCM is composed of a single convolutional layer designed to imitate the learnable color correction matrix. The \textit{simRAW-T sRGB} refers to RGB images generated offline using the teacher model. During training, pairs of sRGB images from the BDD100K dataset \cite{yu2020bdd100k} and simRAW-T sRGB images are used to train LCCM. The \textit{simRAW-S sRGB} refers to RGB images generated by the LCCM, and \textit{simRAW-S} represents Bayer pattern RAW images, similar to \textit{simRAW-T}.}
\label{fig:method}
\end{figure*}

To achieve high-speed RAW reconstruction, we propose an innovative and ultra-lightweight model called the Learnable Color Conversion Matrix (LCCM). Unlike complex inverse ISP models \cite{xing2021invertible, zamir2020cycleisp, kim2023paramisp}, LCCM utilizes only a single convolutional layer to approximate the mapping from sRGB images to RAW images. The parameters of the color conversion matrix are learnable during the training process. Inspired by knowledge distillation \cite{hinton2015distilling}, we use the unpaired RAW reconstruction method Unpaired-CycleR2R \cite{li2024efficient} as the teacher model, while the LCCM serves as the student model. The main framework of our RAW reconstruction method is illustrated in Figure \ref{fig:method}.

\subsection{Unpaired RAW Reconstruction Teacher Model}
\label{sec:teacher}
Modern ISP systems in cameras typically consist of five major steps to convert RAW input into the corresponding sRGB output, as illustrated in Figure \ref{fig:method} (a): demosaic \(f_{dm}\), auto white balance \(f_{wb}\), brightness adjustment \(f_{ba}\), color correction \(f_{cc}\), and gamma correction \(f_{gc}\). The overall conversion performed by the camera ISP system can be defined as a composite function \(f_{ISP}\) composed of multiple invertible and tractable functions as follows \cite{li2024efficient}:
\begin{equation}
f_{\mathrm{ISP}}=f_{gc} \circ f_{cc} \circ f_{ba} \circ f_{wb} \circ f_{dm}
\end{equation}
\begin{equation}
\boldsymbol{y}=f_{\mathrm{ISP}} \circ \boldsymbol{x}
\end{equation}
where \(x\) denotes a RAW image, \(y\) denotes an sRGB image, and $\circ$ denotes the function composition operation. The inverse ISP function \(g_{InvISP}\) that mirrors the ISP function \(f_{ISP}\), can be expressed as \cite{li2024efficient}:
\begin{equation}
g_{\mathrm{InvISP}}=g_{d m} \circ g_{w b} \circ g_{b a} \circ g_{c c} \circ g_{g c}
\end{equation}
\begin{equation}
\boldsymbol{x}=g_{\mathrm{InvISP}} \circ \boldsymbol{y}
\end{equation}
where \(g = f^{-1}\) assumed for simplicity.
As shown in Figure \ref{fig:method} (a), the teacher model Unpaired-CycleR2R \cite{li2024efficient} is designed based on a reversible, modular paradigm. It employs unpaired sRGB and RAW images to train both an sRGB-to-RAW inverse ISP for RAW reconstruction and a RAW-to-sRGB ISP. This unsupervised learning approach based on unpaired samples is highly applicable and practical for converting existing sRGB images into RAW format.

\subsection{Learnable Color Correction Matrix}
\label{sec:ccm}
In a camera ISP system, a 3x3 color correction matrix is utilized for color correction and adjustment, ensuring that the colors are restored to match the human visual system. Our proposed method replaces the complex ISP system with a single learnable color correction matrix to learn the mapping from sRGB images to RAW images. The color correction matrix \(C\) can be defined as below:
\begin{equation}
X=Y \bullet\left[\begin{array}{lll}
C_{11} & C_{12} & C_{13} \\
C_{21} & C_{22} & C_{31} \\
C_{31} & C_{32} & C_{33}
\end{array}\right]
\end{equation}

\noindent where \(X\) denotes a RAW image matrix, \(Y\) denotes an sRGB image matrix, the \textbullet{} denotes the dot product of matrix. The calculation process of the convolutional layer can be represented by the following formula:
\begin{equation}
x_{i j}=\sum_{u=1}^{\mathrm{U}} \sum_{v=1}^{\mathrm{V}} w_{u v} y_{i+u-1, j+v-1}
\end{equation}
where \(x\) denotes a RAW image, \(y\) denotes an sRGB image, \(w\) denotes the convolutional kernel weight parameters, \(i\) and \(j\) denote the pixel positions, \(U\) and \(V\) denote the size of the convolutional kernel. A convolutional layer with a kernel size of 1, stride of 1, input and output channels of 3, and padding of 0, contains 9 weight parameters and 3 bias parameters. We innovatively use the 9 weight parameters of the convolutional layer to simulate the 9 parameters of the color correction matrix. Thus, a color correction matrix with trainable parameters is created, allowing it to learn and approximate the mapping from sRGB images to RAW images. Additionally, the bias parameters are used to fine-tune the color conversion matrix. We calculate the mean squared error between the simRAW images generated by LCCM and those generated by the teacher model Unpaired-CycleR2R \cite{li2024efficient}. The loss function is formulated as follows:
\begin{equation}
L_{M S E}=\frac{1}{n} \sum_{i=1}^n\left(x_{\text {simRAW-S }}-x_{\text {simRAW-T }}\right)^2
\end{equation}
where $x_{\text {simRAW-S }}$ denotes the simRAW images generated by our LCCM, $x_{\text {simRAW-T }}$ denotes the simRAW images generated by the teacher model Unpaired-CycleR2R \cite{li2024efficient}, and $i$ denotes each pixel.

\begin{figure*}
\centering%% For centre alignment of image.
\includegraphics[width=\linewidth]{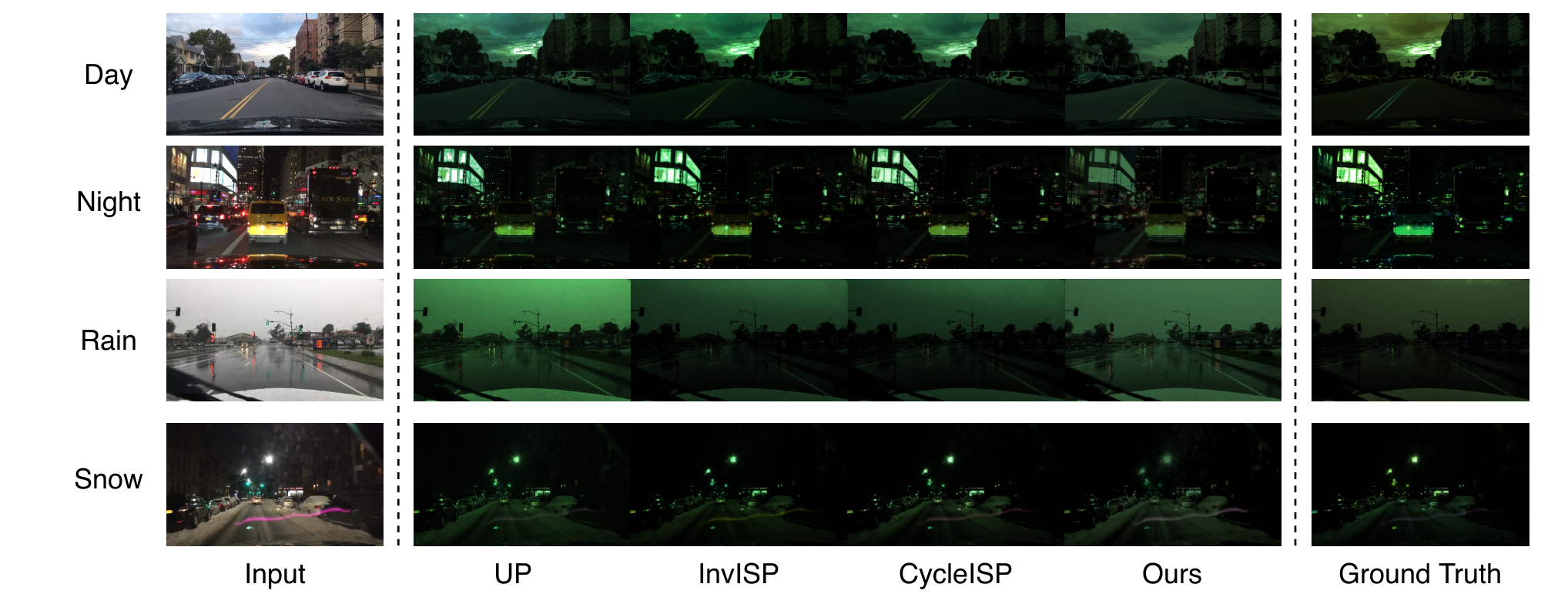}
    \caption{Qualitative comparison of RAW reconstruction results for the asi 294mcpro camera style.}
\label{fig:vis}
\end{figure*}

\section{Experiments}
\subsection{Experiment Settings}
{\bf Datasets.} We evaluate our proposed method on the multiRAW dataset \cite{li2024efficient} and the BDD100K dataset \cite{yu2020bdd100k}. The BDD100K dataset \cite{yu2020bdd100k} is a large-scale dataset contains 100k sRGB images collected in diverse scenes (city streets, residential areas, and highways). The multiRAW dataset \cite{li2024efficient} contains driving scene images covering various conditions (daytime, nighttime, tunnels and rainy weather).

\noindent{\bf Training Details for RAW reconstruction.} Since the teacher model Unpaired-CycleR2R \cite{li2024efficient} is capable of converting sRGB images to RAW images for any camera style, we utilized this open-source model to reconstruct sRGB images from the BDD100K dataset \cite{yu2020bdd100k} into simRAW-T images. This process provides us with sRGB and RAW image pairs, which are then used to train the LCCM. Subsequently, simRAW-S images are generated using the student model LCCM. Since the mapping from sRGB images to RAW images is a pixel-level task, we train the model using the original size of input sRGB images with a batch size of 1. An Adam optimizer with a learning rate of 0.001 is used for 100 training epochs. The quality of the generated simRAW images is evaluated using PSNR and SSIM.

\noindent{\bf Training Details for RAW-domain object detection.} The generated simRAW images can be used to train RAW-domain models for various tasks. We focus on high-level object detection task and select YOLOv3 \cite{redmon2016you} due to its widespread use. Our method can be easily extended to other object detectors. The RAW images are resized to a fixed resolution of 1920×1920. Demosaic and gamma correction are applied in the RAW image processing pipeline as outlined in \cite{li2024efficient}. Data augmentation techniques, including random scaling, cropping, and color jitter, are employed. We use an SGD optimizer with a batch size of 4, a learning rate of 0.02, momentum of 0.9, and weight decay of $10^{-4}$. A linear learning rate adjustment strategy is implemented during training. We train the detector for 30 epochs following \cite{li2024efficient}. Our experiments consist of two stages: pretraining and fine-tuning. In the pretraining stage, the RAW-domain YOLOv3 detector is trained on the simRAW-T and simRAW-S datasets, respectively. In the fine-tuning stage, the pretrained model is used to fine-tune the RAW-domain YOLOv3 detector on the multiRAW dataset \cite{li2024efficient}. 

\subsection{Main Results of RAW Reconstruction}
\begin{figure*}
\centering%% For centre alignment of image.
\includegraphics[width=\linewidth]{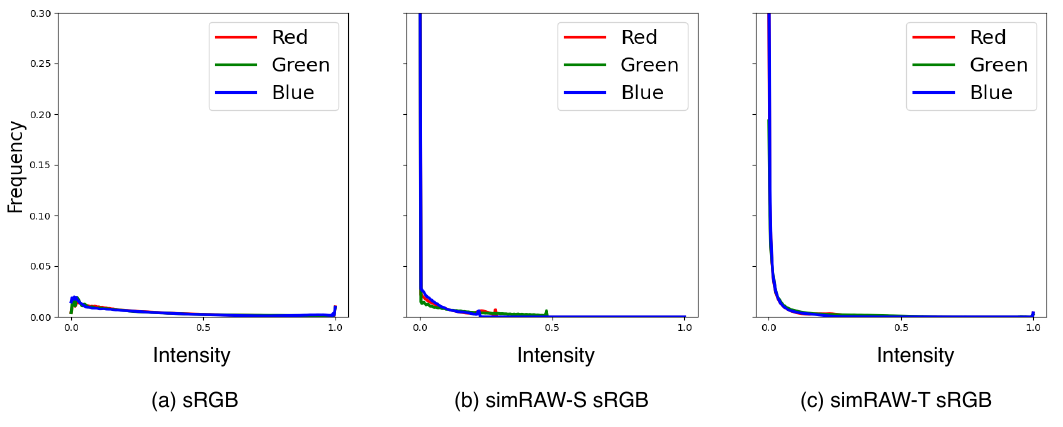}
\caption{Histograms for RGB channel of sRGB images from the BDD100K dataset \cite{yu2020bdd100k}, simRAW-S sRGB and simRAW-T sRGB images for the huawei P30pro camera style. Frequency denotes the occurrence rate of each pixel. Intensity denotes the brightness level of color.}
\label{fig:histrgam}
\end{figure*}

We randomly select 1000 sRGB images from the BDD100K \cite{yu2020bdd100k} training set and 1000 RAW counterparts from simRAW-T images as paired training samples, with 100 of each type as testing samples. As shown in Table \ref{tab:ccm}, the proposed LCCM demonstrates significantly lower parameter and computational complexity, allowing it to efficiently complete both training and inference and contribute to the rapid expansion of RAW image datasets. In terms of PSNR, LCCM surpasses the traditional non-learning-based method UP \cite{brooks2019unprocessing} by 4.03 points, achieves nearly the same point as the learning-based method InvISP \cite{xing2021invertible}, and is only 2.08 points behind CycleISP \cite{zamir2020cycleisp}. 
Figure \ref{fig:vis} illustrates that our proposed LCCM, composed of just a single convolutional layer, can effectively achieve RAW reconstruction under varying weather and lighting conditions. 

Figure \ref{fig:histrgam} presents the histograms for each color channel of the sRGB images from BDD100K \cite{yu2020bdd100k}, simRAW-S sRGB and simRAW-T sRGB images. It is noteworthy that the intensity distributions of simRAW-T sRGB and simRAW-S sRGB images are generally similar, demonstrating that our LCCM can successfully simulate the mapping from sRGB images to RAW images. However, it is observed that the peak distributions of the three color channels in simRAW-S sRGB images differ from those in simRAW-T sRGB images. We attribute this to the limitation of using a single linear convolutional layer, which cannot fully approximate the nonlinear complex inverse ISP. 
While it is expected that a model using only one convolutional layer for RAW reconstruction may not outperform more complex networks, it is important to note that the RAW images generated by this simple network have comparable effects when used to pretrain RAW-domain detectors, as compared to complex inverse ISP methods. This will be further discussed in Sec \ref{sec:det result}.

\begin{table}
\begin{center}
\begin{tabular}{c|c|c|c}
\hline
Method   & PSNR↑          & Param(M)↓    & GFLOP↓    \\ \hline
UP\cite{brooks2019unprocessing}       & 26.67          & -         & -         \\
InvISP\cite{xing2021invertible}   & 30.89          & 1.41      & 577       \\
CycleISP\cite{zamir2020cycleisp} & \textbf{32.78} & 3.14      & 1432      \\ \hline
Ours     & 30.70          & \textbf{1.20 $\times \mathbf{10^{-5}}$}
 & \textbf{3.68 $\times \mathbf{10^{-3}}$}  \\ \hline
\end{tabular}
\end{center}
\caption{Quantitative comparison of RAW reconstruction results for the asi 294mcpro camera style.}
\label{tab:ccm}
\end{table}

\begin{table}
\begin{center}
\begin{tabular}{c|c|c|c}
\hline
Pretraining & Fine-tuning                     & AP↑           & Recall↑       \\ \hline
×           & \multirow{3}{*}{iphone XSmax}  & 76.3          & 89.9          \\ \cline{1-1} \cline{3-4} 
simRAW-T    &                                & \textbf{82.3} & \textbf{92.0} \\ \cline{1-1} \cline{3-4} 
simRAW-S    &                                & 82.1          & 91.3          \\ \hline
×           & \multirow{3}{*}{huawei P30pro} & 67.7          & 81.2          \\ \cline{1-1} \cline{3-4} 
simRAW-T    &                                & \textbf{71.9} & 83.4          \\ \cline{1-1} \cline{3-4} 
simRAW-S    &                                & 71.8          & \textbf{83.7} \\ \hline
×           & \multirow{3}{*}{asi 294mcpro}  & 72.4          & 85.8          \\ \cline{1-1} \cline{3-4} 
simRAW-T    &                                & \textbf{75.8} & 87.0            \\ \cline{1-1} \cline{3-4} 
simRAW-S    &                                & 75.6          & \textbf{87.3} \\ \hline
\end{tabular}
\end{center}
    \caption{Quantitative results of the RAW-domain YOLOv3 detector pretrained on simRAW images. The × denotes training the detector from scratch. To mitigate the impact of the countries where the dataset was captured, Average Precision (AP) and Recall are evaluated only on the vehicle category.}
\label{tab:det}
\end{table}
\subsection{Main Results of RAW-domain Object Detection}
\label{sec:det result}

The purpose of this study is to explore an ultra-simple and highly efficient method for RAW reconstruction, aimed at expanding RAW image datasets to support research in RAW-domain perception tasks for autonomous driving. We focus on the performance of a RAW-domain detector pretrained using simRAW images to verify the feasibility of our proposed LCCM method. The quantitative experimental results are presented in Table \ref{tab:det}. 
First, We randomly initialize the RAW-domain YOLOv3 model and train it from scratch. Then, we pretrain the RAW-domain YOLOv3 detector using simRAW-T and simRAW-S images, respectively, to obtain a pretrained model. This pretrained model is then used to initialize the model parameters and fine-tune the RAW-domain YOLOv3 detector on the multiRAW dataset \cite{li2024efficient}. 

As shown in Table \ref{tab:det}, detection accuracy improves consistently across different camera styles. The results indicate that fine-tuning a simRAW-pretrained YOLOv3 detector outperforms training the model from scratch. Compared to the baseline without pretraining, using LCCM-generated simRAW-S images for pretraining results in an improvement of 5.8 points (iPhone XSmax), 4.1 points (huawei P30Pro) and 3.2 points (asi 294mcpro) in metric AP.
We emphasize that the performance gains obtained by pretraining with simRAW-T images and simRAW-S images are nearly identical. The experimental results reveal the feasibility and significant potential of our proposed LCCM, which is highly efficient, substantially reducing time and computational resources for RAW reconstruction.

\begin{figure*}
\centering%% For centre alignment of image.
\includegraphics[width=\linewidth]{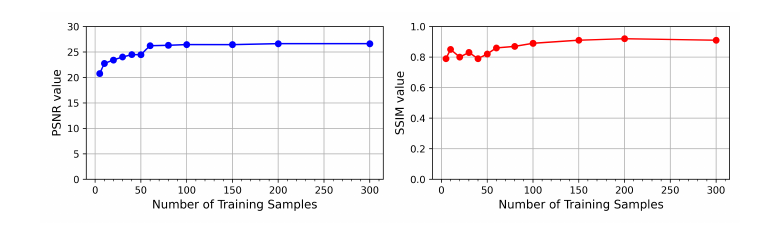}
\caption{Ablation study on the number of training samples for the iphone XSmax camera style.}
\label{fig:psnr-ssim}
\end{figure*}

\subsection{Ablation Studies}
The mapping from sRGB images to RAW images is a pixel-level task, so each high-resolution image provides a large number of pixel samples for training the learnable color conversion matrix. In this section, we conduct experiments to investigate the impact of training sample quantity on the quality of RAW reconstruction images. Our proposed LCCM is particularly notable for its ability to train the color correction matrix using only a few dozen pairs of sRGB and RAW images. As shown in Figure \ref{fig:psnr-ssim}, performance saturation occurs with approximately 100 training samples, This indicates that with just a small set of sRGB and RAW paired images from a specific camera, LCCM can effectively learn the sRGB-to-RAW mapping relationship. These results underscore the efficiency of our proposed LCCM.

\section{Conclusion}
In this paper, we present an innovative inverse ISP method for RAW reconstruction. Our approach leverages a learnable color correction matrix to approximate the mapping from sRGB images to RAW images. The proposed model is exceptionally simple and lightweight, consisting of only a single convolutional layer. Experimental results show that when used to pretrain a RAW-domain object detector, the simRAW images generated by our proposed model achieve performance gains equivalent to those generated by more complex inverse ISP models. It demonstrates the effectiveness and feasibility of our method. We hope that our work will contribute to the rapid expansion of RAW datasets to support research in autonomous driving tasks and inspire new insights into the design of efficient inverse ISP methods.

\section*{Acknowledgements}
This work was supported in part by the National High Quality Program under Grant TC220H07D, in part by the National Key R\&D Program of China under Grant 2022YFB2902002, in part by the Innovation Program of Shanghai Municipal Science and Technology Commission under Grant 20511106603, and in part by Foshan Science and Technology Innovation Team Project under Grant FS0AAKJ919-4402-0060. 

\bibliography{egbib}
\end{document}